\documentclass[prd,aps,twocolumn,10pt,letterpaper,superscriptaddress]{revtex4-2}
\usepackage[utf8]{inputenc}
\usepackage[english]{babel}
\usepackage{bm}
\usepackage{graphicx}
\usepackage{amsmath}
\usepackage{amssymb}
\usepackage{amsfonts}
\usepackage{xcolor}
\usepackage{braket}
\usepackage{hyperref}
\hypersetup{
    pdfstartview={FitH},    % fits the width of the page to the window
    colorlinks=true,       % false: boxed links; true: colored links
    linkcolor=blue,          % color of internal links
    citecolor=blue,        % color of links to bibliography
    filecolor=blue,      % color of file links
    urlcolor=blue           % color of external links
}

\begin{document}

\title{Angular and energy distributions of positrons created in subcritical and supercritical slow collisions of heavy nuclei}

\author{N. K. Dulaev}
\email{st069071@student.spbu.ru}
\affiliation{Department of Physics,
St.~Petersburg State University, 7-9 Universitetskaya nab., St.~Petersburg 
199034, Russia}
\affiliation{Petersburg Nuclear Physics Institute named by B.~P.~Konstantinov of National Research Center ''Kurchatov Institute'', Orlova roscha 1, 188300 Gatchina, Leningrad region$,$ Russia}

\author{D. A. Telnov}
\email{d.telnov@spbu.ru}
\affiliation{Department of Physics,
St.~Petersburg State University, 7-9 Universitetskaya nab., St.~Petersburg 
199034, Russia}

\author{V. M. Shabaev}
\email{v.shabaev@spbu.ru}
\affiliation{Department of Physics,
St.~Petersburg State University, 7-9 Universitetskaya nab., St.~Petersburg 
199034, Russia}
\affiliation{Petersburg Nuclear Physics Institute named by B.~P.~Konstantinov of National Research Center ''Kurchatov Institute'', Orlova roscha 1, 188300 Gatchina, Leningrad region$,$ Russia}

\author{Y.~S.~Kozhedub}
\affiliation{Department of Physics,
St.~Petersburg State University, 7-9 Universitetskaya nab., St.~Petersburg 
199034, Russia}

\author{I. A. Maltsev}
\affiliation{Department of Physics,
St.~Petersburg State University, 7-9 Universitetskaya nab., St.~Petersburg 
199034, Russia}

\author{R. V. Popov}
\affiliation{Department of Physics,
St.~Petersburg State University, 7-9 Universitetskaya nab., St.~Petersburg 
199034, Russia}
\affiliation{Petersburg Nuclear Physics Institute named by B.~P.~Konstantinov of National Research Center ''Kurchatov Institute'', Orlova roscha 1, 188300 Gatchina, Leningrad region$,$ Russia}

\author{I. I. Tupitsyn}
\affiliation{Department of Physics,
St.~Petersburg State University, 7-9 Universitetskaya nab., St.~Petersburg 
199034, Russia}

\begin{abstract}
Positron creation probabilities as well as energy and angular distributions of outgoing positrons in slow collisions of two identical heavy nuclei are obtained within the two-center approach beyond the monopole approximation. The time-dependent Dirac equation for positron wave functions is solved with the help of the generalized pseudospectral method in modified prolate spheroidal coordinates adapted for variable internuclear separation. Depending on the nuclear charge, the results are obtained for both subcritical and supercritical regimes of the positron creation. The signatures of transition to the supercritical regime in the total positron creation probabilities and energy spectra are discussed. The angular distributions of emitted positrons demonstrate a high degree of isotropy.
\end{abstract}

\maketitle

\section{Introduction}

The electron-positron pair creation in the nonperturbative regime of quantum electrodynamics (QED) in the presence of strong electromagnetic fields is a subject of intense theoretical studies (see, e.g., reviews \cite{Ehlotzky_2009_Fundamental, Ruffini_2010_Electron, Piazza_2012_Extremly, Fedotov_2023_Advances}). The experimental verification of existence of such processes is of great difficulty since the required critical field strength is extremely high ~--- at the level of $10^{16}$ V/cm. Although the peak field strength achieved with laser technologies has ever increased over the past decades, it is still impossible to obtain supercritical values. Another possibility to achieve supercritical fields arises in the physics of heavy nuclei and heavy-ion collisions. 

In a pioneering paper \cite{Pomeranchuk_1945} it was shown that the $1s$ level of a hydrogenlike ion with an extended nucleus gradually decreases with increasing $Z$ and at a certain critical $Z$, $Z=Z_{\mathrm{cr}}$, reaches the negative-energy continuum.
In papers of Soviet and German physicists \cite{Gershtein_1969, Pieper_1969_Interior, Popov_1970_1, *Popov_1970_2, *Popov_1970_3, *Popov_1971_1, Zeldovich_1971_Electronic, Muller_1972_Solution, *Muller_1972_Electron, Mur_1976, Popov_1976, Muller_1976_Positron, Reinhardt_1977_Quantum, Soff_1977_Shakeoff, Rafelski_1978_Fermions, Greiner_1985_Quantum} it was shown that the diving of an initially empty $1s$ state into the negative-energy electron continuum can result in spontaneous emission of positrons. In this process the originally neutral vacuum decays into the charged vacuum and two positrons. For the $1s$ state $Z_{\mathrm{cr}} \approx 173$, and since there is no experimental means to produce such a heavy nucleus in the near future, low-energy heavy ion collisions got major attention. If during the collision process two heavy nuclei with the charge numbers $Z_1 + Z_2 > Z_{\mathrm{cr}}$ get sufficiently close to each other, for a short period of time the quasimolecular $1s\sigma_g$ state dives into the negative-energy continuum as a resonance, resulting in the creation of electron-positron pairs, and the positrons can escape the nuclei as free particles. 

The first calculations of spontaneous vacuum decay were carried out in the static approximation \cite{Popov_1973, Peitz_1973_Autoionization, Popov_1979}, which does not take into account the dynamical pair production mechanism, arising from the variation of the potential due to the nuclei motion. The spontaneous and dynamical pair creation channels were examined in papers by the Frankfurt group (see, e.g., \cite{Smith_1974_Induced, Reinhardt_1981_Theory, Muller_1988_Positron, Rafelski_1978_Fermions, Greiner_1985_Quantum, Bosch_1986_Positron, Muller_1994_Electron, Reinhardt_2005_Supercritical, Rafelski_2017_Probing}). However, the experiments performed many years ago at GSI (Darmstadt) did now show any signature of the spontaneous pair production (see, e.g., Refs. \cite{Greiner_1985_Quantum, Muller_1994_Electron, Reinhardt_2005_Supercritical, Rafelski_2017_Probing} and references therein). This was mainly due to a strong masking of the spontaneous pair-creation channel by the dynamical one.
As a result, the Frankfurt group concluded that experimental verification of the spontaneous positron creation is only possible if the nuclei ``stick'' to each other during the collision process due to nuclear forces, enhancing the spontaneous channel  \cite{Reinhardt_2005_Supercritical}. However, since there is no experimental evidence of the nuclear ``sticking'' to date, other approaches need to be considered.

In the last two decades, theoretical interest in the spontaneous pair creation has risen again. 
The superctritical resonance was studied in Refs. \cite{Ackad_2007_Numerical, Godunov_2017_Resonances, Ackad_2007_Supercritical, Marsman_2011_Calculation, Maltsev_2020_Calculation}. In Refs.~\cite{Grashin_2022_Vacuum, Krasnov_2022_Non}, the effects of the QED-vacuum polarization in the supercritical Coulomb field have been evaluated. Pair creation in heavy-ion collisions in dynamic framework was targeted in the monopole \cite{Ackad_2008_Calculation, Maltsev_2015_Electron, Bondarev_2015_Positron} and beyond the monopole approximation \cite{Maltsev_2017_Pair, Maltsev_2018_Electron, Popov_2018_One}. In Ref. \cite{Voskresensky_2021_Electron}, the instability of electron-positron vacuum in the relativistic semiclassical approach was examined.

In Refs. \cite{Maltsev_2019_How, Popov_2020_How}, a new method to study supercritical regime signatures was proposed. The approach is to examine collisions with nuclei moving along trajectories with a fixed minimal internuclear distance $R_{\mathrm{min}}$ and different energy parameter $\varepsilon=E/E_0$, $\varepsilon \geq 1$, where $E$ is the collision energy and $E_0$ is the head-on collision energy. Moving along these trajectories, the nuclei with charges $Z_1 + Z_2 > Z_{\mathrm{cr}}$ produce the same field strength but spend different time in the supercritical regime depending on $\varepsilon$. With increasing $\varepsilon$, the time of the supercritical regime and the contribution of the spontaneous pair creation channel decrease. On the contrary, the dynamical channel contribution increases for greater collision energies $E$. Thus, the increase in the pair creation probability with decreasing $\varepsilon$ should be referred to the spontaneous regime. In Refs. \cite{Maltsev_2019_How, Popov_2020_How}, the signatures of the transition to the supercritical regime were found in the monopole approximation, and in Ref. \cite{Popov_2023_Spontaneous} the analogous results were obtained in the calculations beyond the monopole-approximation framework. 

With the new possibilities, which are anticipated at the upcoming experimental facilities in Germany (GSI/FAIR) \cite{Gumberidze_2009_X, Lestinsky_2016_Physics}, China (HIAF) \cite{Ma_2017_HIAF}, and Russia \cite{Akopian_2015_Layout}, further theoretical investigations of the pair creation in low-energy heavy nuclei collisions are needed. Among other characteristics, the energy spectra of outgoing positrons with angular resolution are of great interest. The pair-production differential probability with respect to the energy and angles would provide valuable information for possible experimental setup construction. If the positron angular distributions have a high degree of isotropy, fewer detectors would be needed to conduct the experiment. Previously, the problem of the angular distributions was targeted in Refs. \cite{Popov_1973, Popov_1979} in the framework of static approximation.

The present work aims to study the angular-energy spectra of positrons in low-energy heavy-ion collisions. The generalized pseudospectral method is used to solve the time-dependent Dirac equation in modified prolate spheroidal coordinates. The rotational term in the time-dependent Dirac equation appearing in the rotating molecular reference frame (see, e.g., Ref. \cite{Muller_1976_The}) as well as the magnetic field of the nuclei are omitted. In Refs. \cite{Betz_1976_Direct, Soff_1979_Electrons, Reinhardt_1980_Dynamical, Soff_1988_Positron} it was argued that the contribution of these effects to the total pair creation probability and the energy distributions is negligible. 
%Besides, the calculations including the rotational term would require much more computing power to process.%
The plane wave decomposition approach is applied to obtain the energy spectra of positrons with angular resolution. The supercritical signatures in the angle-integrated as well as angle-resolved energy distributions are studied. The angular anisotropy of the positron distributions is examined. 

The paper is organized as follows. In Sec.~\ref{methods}, the theoretical methods applied to solve the time-dependent Dirac equation and calculate the positron spectra are described. In Sec.~\ref{results}, the results for the angle-integrated and angle-resolved energy spectra of outgoing positrons are presented and discussed.

Atomic units ($\hbar=|e|=m_{e}=1$) are used throughout the paper unless specified otherwise.

\section{Methods}\label{methods}
\subsection{Time-dependent Dirac equation for a one-positron diatomic quasimolecular system}

\begin{figure}
   \includegraphics[width=0.8\columnwidth]{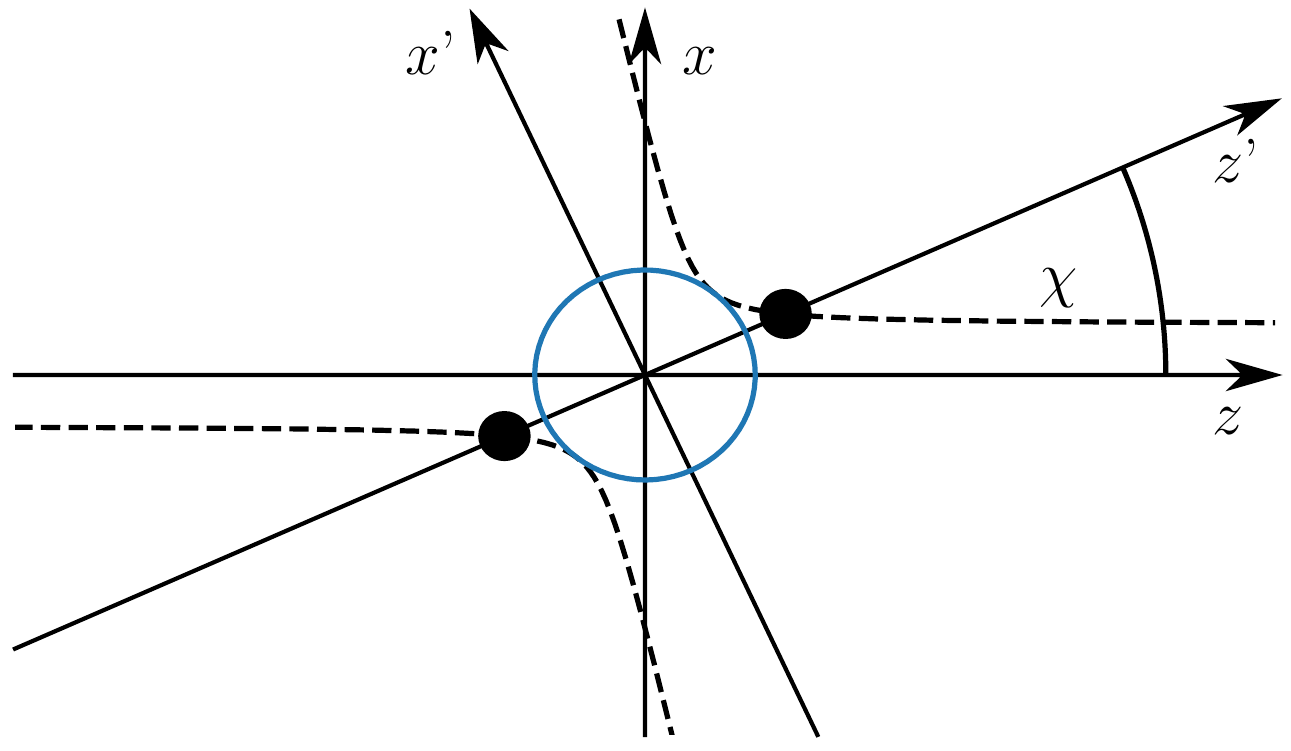}
    \caption{The principal scheme of the nuclei collision. Here, $z-x$ axes correspond to the fixed initial large internuclear separation frame of reference, $z'-x'$ axes correspond to the rotating molecular frame of reference, filled black circles and dotted lines depict the nuclei and their trajectories, respectively, and $\chi$ is the internuclear axis rotation angle. The blue circle is the closest nuclei approach distance.}
    \label{fig:trj}
\end{figure}
We study a collision of two identical bare nuclei and use Dirac's hole picture based on the Dirac equation for positrons \cite{Greiner_1985_Quantum,Godunov_2017_Resonances}. In this approach, the lower Dirac continuum states as well as discrete states, detached from the lower continuum to the gap between the lower and upper continua, are considered occupied by positrons and form the Dirac vacuum. It should be emphasized that the discrete states in the gap between the lower and upper continua are not the states of a real positron, which cannot be bound in a repulsive nuclear potential. A positron becomes observable if a transition is made from such states to the upper continuum. At the same time, a hole in the lower continuum or in the discrete state is created, which is described as an electron in a continuum or bound state, respectively. The electron-positron pair creation can be caused by absorption of the energy from external time-dependent fields or may occur spontaneously if the parameters of the nuclear system, such as internuclear separation, make it possible that the energy of some positron vacuum state get in the upper continuum.

In the center-of-mass frame of reference, the time-dependent Dirac equation (TDDE) for a positron subject to the static electric field of the nuclei reads as
\begin{equation}
    i \frac{\partial}{\partial t} \Psi(\bm{r}, t) = H\Psi(\bm{r},t), \label{eq:tdde}
\end{equation}
where $\Psi(\bm{r},t)$ is a four-component wave function of the positron, and the Hamiltonian $H$ takes the form
\begin{equation}
\begin{split}
    H &= c(\bm{\alpha} \cdot \bm{p}) + c^2 \beta\\ 
    &+ U_{n}(|\bm{r}-\bm{a}(t)|) + U_{n}(|\bm{r}+\bm{a}(t)|).
\end{split}
    \label{eq:hamiltonian}
\end{equation}
Here, $c$ is the speed of light, $\bm{p}$ is the momentum operator, $\bm{\alpha}$ and $\beta$ are the Dirac matrices, and the vector $\bm{a}(t)$ is directed along the instantaneous internuclear axis with its length equal to one half of the internuclear distance $R(t)$:
\begin{equation}
    \bm{a} = a(t)A(t)\bm{e}_z, \quad a(t) = \frac{1}{2} R(t). \label{eq:scaling_fact}
\end{equation}
Here, $\bm{e}_z$ is the unit vector along the $z$ axis, $A(t)$ is an orthogonal $3\times 3$ matrix describing the rotation of the internuclear axis during the collision. Assuming the rotation of the internuclear axis in the $z - x$ plane (the collision process scheme is depicted in Fig. \ref{fig:trj}), one can write the matrix $A(t)$ as
\begin{equation}
    A(t) = \begin{pmatrix}
        \cos{\chi} & 0 & \sin{\chi} \\
        0 & 1 & 0 \\
        - \sin{\chi} & 0 & \cos{\chi}
        \end{pmatrix}.
\end{equation}
The rotation angle $\chi$ depends on time, with the original direction of the internuclear axis along the $z$ axis ($\chi \rightarrow 0$ as $t \rightarrow -\infty$).

The potential $U_{n}(r)$ in Eq.~(\ref{eq:hamiltonian}) is spherically symmetric and represents the static electric potential of the nucleus within the extended nucleus model:
\begin{equation}
    U_{n}(r) = \int d^{3} r' \frac{\rho_{n}(r')}{|\bm{r} - \bm{r}'|},\label{eq:diatomic_pot}
\end{equation}
where $\rho_n(r)$ is the nuclear charge distribution function. 

With the help of the following unitary transformation:
\begin{equation}
    \Psi(\bm{r}, t) = \exp{\left[ -i\chi J_y \right]} \Psi^{(r)}(\bm{r},t).\label{eq:wv_in_rot}
\end{equation}
we make a transition to the rotated frame of reference. Here $J_y$ is the operator representing the total angular momentum projection onto the $y$ axis:
\begin{equation}
    J_y = L_y + S_y,
\end{equation}
where $L_y$ is the orbital angular momentum projection operator and $S_y$ is the spin angular momentum projection operator. 
Upon substitution of the wave function (\ref{eq:wv_in_rot}) in the equation (\ref{eq:tdde}) one obtains:
\begin{equation}
\begin{split}
    &i \frac{\partial \Psi^{(r)}(\bm{r}, t)}{\partial t} = \Big[c(\bm{\alpha} \cdot \bm{p}) + c^2 \beta - \dot{\chi} J_y\\
    &+U_{n}(|\bm{r}-a(t)\bm{e}_{z}|) + U_{n}(|\bm{r}+a(t)\bm{e}_{z}|)
\Big] \Psi^{(r)}(\bm{r}, t).
\end{split}
\label{eq:eq_rotated}
\end{equation}
We note that $\Psi^{(r)}(\bm{r}, t)$ is just an auxiliary wave function defined by Eq.~(\ref{eq:wv_in_rot}), and Eq.~(\ref{eq:eq_rotated}) differs
from the equation for the true wave function in the relativistic rotating frame of reference; for the details, see Ref.~\cite{Muller_1976_The}. 

In Eq.~(\ref{eq:eq_rotated}) the nuclei are always on the Cartesian $z$ axis but the rotational coupling operator $-\dot{\chi} J_y$ appears as an additional term in the Hamiltonian. The next unitary transformation, \begin{equation}
    \Psi^{(r)}=\begin{pmatrix}
        1&0&0&0\\
        0&\exp(i\varphi)&0&0 \\
        0&0&i \\
        0&0&0&i\exp(i\varphi)
    \end{pmatrix}\psi , 
\end{equation}
where the angle $\varphi$ describes the rotation about the internuclear axis, results in a time-dependent equation for the wave function $\psi(t)$,
\begin{equation}
    i \frac{\partial}{\partial t} \psi(t) = \left[ \mathsf{H} - \dot{\chi} \mathsf{J}_y \right] \psi(t). \label{eq:tdde_aft_trans}
\end{equation}
where the new Hamiltonian $ \mathsf{H}$ reads as: 
\begin{equation}
\begin{split}
    &\mathsf{H} = c^2 
    \begin{Vmatrix}
        1_{2} & 0_{2}  \\
        0_{2} & -1_{2}
    \end{Vmatrix}
    + c
    \begin{Vmatrix}
        0_{2} & B  \\
        B^\dag & 0_{2}
    \end{Vmatrix}
    + c
    \begin{Vmatrix}
        0_{2} & D  \\
        D^\dag & 0_{2}
    \end{Vmatrix}\\
    &+ \Big[U_{n}(|\bm{r}-a(t)\bm{e}_{z}|) + U_{n}(|\bm{r}+a(t)\bm{e}_{z}|)\Big]
    \begin{Vmatrix}
        1_{2} & 0_{2}  \\
        0_{2} & 1_{2}
    \end{Vmatrix}, \label{eq:H_tdde_aft_trans}
\end{split}
\end{equation}
and notations $0_{2}$ and $1_{2}$ stand for the zero and unit $2\times2$ matrices, respectively. Using the cylindrical coordinates $\rho$, $z$, and $\varphi$ as generic coordinates to represent the differential operators, the $2 \times 2$ matrices $B$ and $D$
in Eq. (\ref{eq:H_tdde_aft_trans}) can be written as follows:
\begin{equation}
    B = \begin{pmatrix}
        \dfrac{\partial}{\partial z} & \dfrac{\partial}{\partial \rho} + \dfrac{1}{\rho} \\
        \dfrac{\partial}{\partial \rho} & - \dfrac{\partial}{\partial z}
    \end{pmatrix}, \quad
    D = \begin{pmatrix}
       0 & - \dfrac{i}{\rho} \dfrac{\partial}{\partial \varphi} \\
        \dfrac{i}{\rho}  \dfrac{\partial}{\partial \varphi} & 0
    \end{pmatrix}. \label{eq:H_matrices}
\end{equation}
These matrices are anti-Hermitian:
\begin{equation}
    B^\dag = - B, \quad D^\dag = - D.
\end{equation}
The Hamiltonian $\mathsf{H}$ in Eq.~(\ref{eq:H_tdde_aft_trans}) is real-valued except for the term with the matrix $D$. The transformed angular momentum operator $\mathsf{J}_y$ is defined as:
\begin{equation}
    \mathsf{J}_y = \mathsf{L}_y + \mathsf{S}_y, \label{eq:jy}
\end{equation}

\begin{equation}
    \mathsf{L}_y = L_y - \frac{z}{\rho} \sin{\varphi} 
    \begin{pmatrix}
        0 & 0 & 0 & 0 \\
        0 & 1 & 0 & 0 \\
        0 & 0 & 0 & 0 \\
        0 & 0 & 0 & 1 
    \end{pmatrix}, \label{eq:ly}
\end{equation}
\begin{equation}
\begin{split}
    \mathsf{S}_y = &\frac{i}{2} \exp(-i\varphi) 
    \begin{pmatrix}
        0 & 0 & 0 & 0 \\
        1 & 0 & 0 & 0 \\
        0 & 0 & 0 & 0 \\
        0 & 0 & 1 & 0 
    \end{pmatrix}\\
&-\frac{i}{2} \exp(i\varphi) 
    \begin{pmatrix}
        0 & 1 & 0 & 0 \\
        0 & 0 & 0 & 0 \\
        0 & 0 & 0 & 1 \\
        0 & 0 & 0 & 0 
    \end{pmatrix}    .\label{eq:sy}
\end{split}
\end{equation}

\subsection{Modified prolate spheroidal coordinates and time propagation}
Prolate spheroidal coordinates are a natural choice for description of two-center quantum systems. Conventional prolate spheroidal coordinates $\xi$ and $\eta$ \cite{abra1972} are related to cylindrical coordinates as follows:
\begin{equation}
\begin{split}
    \rho &= a \sqrt{(\xi^2 - 1)(1 - \eta^2)}, \quad z  = a \xi \eta\\
    &(1\le\xi<\infty, -1\le\eta\le 1).
\end{split}
\end{equation}
The angle $\varphi$ has the same definition in both coordinate systems. However, conventional prolate spheroidal coordinates are not well suited for numerical calculations of close collisions, when the parameter $a(t)$ may become very small, because the physical volume is determined by the product $a\xi$, and not by $\xi$ alone. That is why we use \textit{modified} prolate spheroidal coordinates for solving the time-dependent equation (\ref{eq:tdde_aft_trans}). Instead of the coordinate $\xi$, we introduce a new coordinate $\lambda$ according to the following definition:
\begin{equation}
    \lambda = a (\xi - 1),
\end{equation}
so the relations between the cylindrical coordinates and modified prolate spheroidal coordinates read as 
\begin{equation}
\begin{split}
    \rho  &= \sqrt{ \left[ (\lambda + a)^2 - a^2 \right] \left[ 1 - \eta^2 \right] }, \\
    z  &= (\lambda + a) \eta\\
    &(0\le\lambda<\infty, -1\le\eta\le 1). \label{eq:mpsc}
\end{split}
\end{equation}
In the limit $a\rightarrow 0$, the modified prolate spheroidal coordinates are smoothly transformed into the spherical coordinates $r$ and $\vartheta$: $\lambda\rightarrow r$, $\eta\rightarrow\cos\vartheta$.

The operators $\mathsf{H}$ in Eq.~(\ref{eq:H_tdde_aft_trans}) and $\mathsf{J}_y$ in Eq.~(\ref{eq:jy}) must be expressed in the modified prolate spheroidal coordinates before solving Eq.~(\ref{eq:tdde_aft_trans}).
Besides that, an additional scaling term is introduced in the equation since the transformation to the modified prolate spheroidal coordinates (\ref{eq:mpsc}) is time-dependent through the parameter $a(t)$:
\begin{equation}
    i \frac{\partial \psi(\lambda, \eta, \varphi, t)}{\partial t} = \left[ \mathsf{H} - \dot{\chi} \mathsf{J}_y - \frac{\dot{a}}{a} \mathsf{S}_a \right] \psi(\lambda, \eta, \varphi, t). \label{eq:final_tdde}
\end{equation}
Expressions of the operators $\mathsf{H}$, $\mathsf{J}_y$, and $\mathsf{S}_a$ in the modified prolate spheroidal coordinates are rather cumbersome; they are not reproduced here and will be published elsewhere. 

As was argued in Refs.~\cite{Betz_1976_Direct,Soff_1979_Electrons,Reinhardt_1980_Dynamical,Soff_1988_Positron}, the influence of the rotational coupling term $-\dot{\chi} \mathsf{J}_y$ in TDDE (\ref{eq:final_tdde}) on the total production of positrons in slow collisions of heavy nuclei is negligible. A question if the rotational coupling and magnetic field of the nuclei have a noticeable effect on the energy and angular distributions of the positrons needs further investigation involving large-scale computations. It may be a subject of future research. In the present work, we adopt an approximation that neglects the rotational coupling term $-\dot{\chi} \mathsf{J}_y$ in TDDE (\ref{eq:final_tdde}), so the positron angular momentum projection on the internuclear axis is conserved. 

To solve the TDDE (\ref{eq:final_tdde}), we employ the generalized pseudospectral (GPS) method in prolate spheroidal coordinates, which was extensively used and discussed previously \cite{Telnov_2007_Ab, Telnov_2009_Effects, Telnov_2018_Multiphoton}. Here, we adapt the method for usage with the modified prolate spheroidal coordinates. To perform time propagation in Eq.~(\ref{eq:final_tdde}), we apply a scheme based on the Crank--Nicolson algorithm \cite{Crank_1947_A},
\begin{multline}
     \bigg[ 1 + \frac{i\Delta t}{2}  \Tilde{\mathsf{H}}(t + \frac{1}{2} \Delta t) \bigg] \psi(t + \Delta t) = \\
     = \bigg[ 1 - \frac{i\Delta t}{2} \Tilde{\mathsf{H}}(t + \frac{1}{2} \Delta t) \bigg] \psi(t), 
\end{multline}
where $\Tilde{\mathsf{H}}(t) = \mathsf{H}(t) - (\dot{a}/a) \, \mathsf{S}_a  - c^2$. With the energy shift of $-c^2$, the onset of the upper positron continuum is placed at zero energy, thus improving the accuracy of the Crank--Nicolson scheme for propagation of the free-positron wave packet. At the end of the propagation process, the positron wave function $\Psi^{(r)}$ is projected onto the upper positron continuum, so the resulting wave packet $\Psi^{(c)}$ represents only free positrons. This wave packet is then used to extract the energy and angular distribution of outgoing positrons.

\subsection{Calculation of positron spectra}
To obtain angular-energy distributions of outgoing positrons after the collision process, we project the free-positron wave packet $\Psi^{(c)}$ onto the plane waves with the momentum $\bm{k}$. Two such wave functions can be constructed, which differ by the spin state of the positron. We make use of the functions with the fixed spin projection on the $z$ axis when the positron is at rest \cite{bere1982}:
\begin{equation}
    \Psi^{(1)} = 
    \begin{pmatrix}
    \sqrt{E_{k} + 2c^2} \\
    0 \\
    \sqrt{E_{k}} \cos{\vartheta_k} \\
    \sqrt{E_{k}} \sin{\vartheta_k} \, \exp(i \varphi_k)
    \end{pmatrix}
    \frac{\exp{[i (\bm{k}\cdot\bm{r}) - iE_{k}t]}}{(2\pi)^{3/2}\sqrt{2E_{k} + 2c^2}},
\end{equation}
\begin{equation}
    \Psi^{(2)} = 
    \begin{pmatrix}
    0 \\
    \sqrt{E_{k} + 2c^2} \, \exp(i \varphi_k)\\
    \sqrt{E_{k}}  \sin{\vartheta_k} \\
    -\sqrt{E_{k}} \cos{\vartheta_k}\, \exp(i \varphi_k)
    \end{pmatrix}
    \frac{\exp{[i (\bm{k}\cdot\bm{r}) - iE_{k}t]}}{(2\pi)^{3/2}\sqrt{2E_{k} + 2c^2}}.
\end{equation}
Here $\vartheta_k$, $\varphi_k$ are the polar and azimuthal angles of the momentum $\bm{k}$, respectively, and $E_{k}$ is the kinetic energy related to the absolute value of the momentum $k$ as
\begin{equation}
    k = \frac{1}{c} \sqrt{E_{k}(E_{k} + 2c^2)}.
\end{equation}
The plane waves $\Psi^{(1)}$ and $\Psi^{(2)}$ are normalized to the delta function in the momentum space, therefore, the differential pair-production probabilities can be written as
\begin{equation}
    \frac{dP^{(1)}}{dE_{k}d\Omega}=\frac{1}{c^3} \sqrt{E_{k}(E_{k}+2c^2)} (E_{k} + c^2) |\braket{\Psi^{(1)}|\Psi^{(c)}}|^2, \label{eq:p1}
\end{equation}
\begin{equation}
    \frac{dP^{(2)}}{dE_{k}d\Omega}=\frac{1}{c^3} \sqrt{E_{k}(E_{k}+2c^2)} (E_{k} + c^2) |\braket{\Psi^{(2)}|\Psi^{(c)}}|^2. \label{eq:p2}
\end{equation}
The positron spin projection on the $z$ axis is conserved in the rest frame of reference but does not have much sense in the laboratory frame where the positron moves. Thus only the sum of the differential probabilities (\ref{eq:p1}) and (\ref{eq:p2}) is meaningful:
\begin{equation}
  \frac{dP}{dE_{k}d\Omega} =  \frac{dP^{(1)}}{dE_{k}d\Omega} +  \frac{dP^{(2)}}{dE_{k}d\Omega}.   \label{eq:distr}
\end{equation}
It provides a distribution of outgoing positrons over energies and angles.

The distribution given by Eq.~(\ref{eq:distr}) is approximate. Strictly speaking, not plane waves but continuum states of the positron in the Coulomb field of the two nuclei should be used to calculate the energy and angular distributions. Construction of such functions with correct asymptotic behavior at large distances is a tough problem. On the other hand, the distribution is calculated after the collision, when the outgoing wave packet $\Psi^{(c)}$ is already far away from the nuclei. In this situation, the differential probability (\ref{eq:distr}) is expected to be a good approximation. We estimate an inaccuracy in the energy distributions at the level of $10$~keV. This is well acceptable since the scale of the energy spectrum is about $500$~keV.

\section{Results}\label{results}
The calculations have been performed for slow collisions of two identical bare nuclei with the charge numbers $Z=83$, $Z=87$, $Z=92$, and $Z=96$ ($Z=Z_{1}=Z_{2}$). The motion of the nuclei is described within the classical mechanics; the law of motion and trajectories are given by the well-known solutions of the Rutherford scattering problem. The positron wave functions are obtained by solving TDDE as described in Sec.~\ref{methods}. For the nuclear charge distribution, we employ the Fermi model \cite{parp1992}, which is widely used and considered reliable \cite{viss1997,shab2013,miro2015}; the root-mean-square radii of the nuclei are taken from the tables \cite{Angeli_2013_Table}.

\begin{figure}
    \includegraphics[width=\columnwidth]{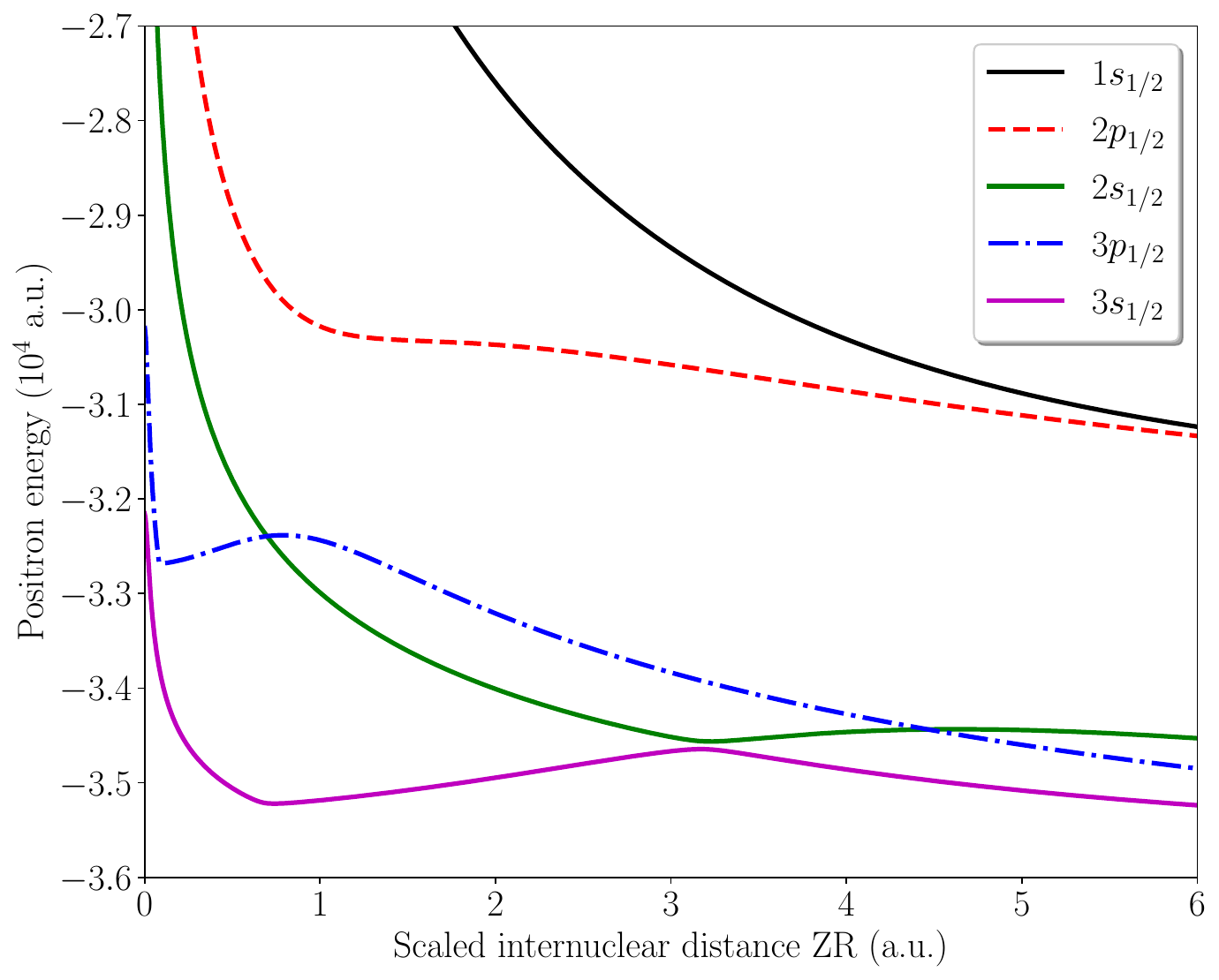}
    \caption{Energies of positron quasimolecular discrete vacuum states for $Z=92$ lying most closely to the upper positron continuum vs internuclear distance. Zero energy is at the onset of the upper positron continuum. The states are labeled by quantum numbers in the united atom limit. 
    }
    \label{fig:energies_Z92}
\end{figure}

The main contribution to the positron production is expected from those discrete vacuum positron states, whose energy levels are shifted close to the onset of the upper positron continuum as the nuclei approach each other  (or even enter this continuum, thus becoming supercritical resonances). Therefore we propagate the discrete quasimolecular states, which can be labeled $1s_{1/2}$, $2p_{1/2}$, $2s_{1/2}$, $3p_{1/2}$, $3s_{1/2}$ in the united atom limit, when the internuclear separation vanishes. In Fig.~\ref{fig:energies_Z92}, we show the energies of these states vs the internuclear distance for $Z=92$; for the other nuclear charge numbers used in the calculations, the pattern of the energy levels is similar. As one can see, transitions to the upper positron continuum must be dominated by the $1s_{1/2}$, $2p_{1/2}$, and $2s_{1/2}$ states, which come most closely to the onset of the upper continuum at small internuclear separations (the $1s_{1/2}$ energy level eventually crosses into the upper continuum at $R\approx 35$~fm). However, there is an avoided crossing between the $2s_{1/2}$ and $3s_{1/2}$ adiabatic curves at $R=3.2/Z$~a.u., and the adiabatic quasimolecular state labeled as $3s_{1/2}$ may have a significant contribution to the positron production as well. Therefore all five discrete positron quasimolecular states mentioned above are included in the time propagation.

The initial and finial internuclear separations in the time propagation are taken at $R_{\mathrm{max}}=5.5/Z$~a.u. At such large internuclear distances, the quasimolecular states under consideration are already close to the separate atoms limit. Avoided crossings between the corresponding adiabatic curves are located at smaller internuclear separations, so possible nonadiabatic transitions, which may affect the positron creation process, are taken into account during the time propagation. Following Refs.~\cite{Popov_2020_How,Popov_2023_Spontaneous}, we consider nuclear trajectories corresponding to the same smallest internuclear separation $R_{\mathrm{min}}$, which is set to $17.5$~fm in all cases. At this $R_{\mathrm{min}}$, the $1s_{1/2}$ positron energy level is deeply in the supercritical regime for $Z=96$ and $Z=92$ while it remains in the subcritical regime for  $Z=87$ and $Z=83$. The $2p_{1/2}$ level briefly enters the upper positron continuum in the vicinity of $R_{\mathrm{min}}$ for $Z=96$. The other energy levels do not reach the onset of the upper positron continuum for all $Z$ used in the calculations. The numerical parameters of the calculations are as follows.
The number of time-propagation steps is $4096$; the spatial grid has $384$ points for the pseudoradial coordinate $\lambda$ and $16$ points for the pseudoangular coordinate $\eta$. The radial box size is equal to $60/Z$ a.u. 

\subsection{Total positron creation probabilities}
\begin{table}
 \caption{Total positron creation probabilities in collisions of two identical nuclei with $R_{\mathrm{min}}=17.5$~fm; $a$ -- present results, $b$ -- Ref.~\cite{Popov_2023_Spontaneous}.}
\begin{ruledtabular}
\begin{tabular}{lcccc}
\multicolumn{1}{c}{Nucleus}&\multicolumn{4}{c}{Scaled collision energy} \\
\cline{2-5}
\multicolumn{1}{c}{charge}&\multicolumn{1}{c}{$\varepsilon=1.0$}&\multicolumn{1}{c}{$\varepsilon=1.1$}&\multicolumn{1}{c}{$\varepsilon=1.2$}&\multicolumn{1}{c}{$\varepsilon=1.3$}\\
\hline
$Z=83^{a}$&$3.88\times 10^{-4}$&$4.39\times 10^{-4}$&$4.83\times 10^{-4}$&$5.24\times 10^{-4}$\\
$Z=87^{a}$&$1.88\times 10^{-3}$&$1.99\times 10^{-3}$&$2.08\times 10^{-3}$&$2.16\times 10^{-3}$\\
$Z=92^{a}$&$1.13\times 10^{-2}$&$1.12\times 10^{-2}$&$1.11\times 10^{-2}$&$1.10\times 10^{-2}$\\
$Z=92^{b}$&$1.20\times 10^{-2}$&$1.20\times 10^{-2}$&$1.20\times 10^{-2}$&$1.20\times 10^{-2}$\\
$Z=96^{a}$&$4.12\times 10^{-2}$&$3.90\times 10^{-2}$&$3.73\times 10^{-2}$&$3.59\times 10^{-2}$\\
$Z=96^{b}$&$4.26\times 10^{-2}$&$4.07\times 10^{-2}$&$3.93\times 10^{-2}$&$3.81\times 10^{-2}$
\end{tabular}
\end{ruledtabular}
\label{tab1}
\end{table}
\begin{figure*}
    \includegraphics[width=\textwidth]{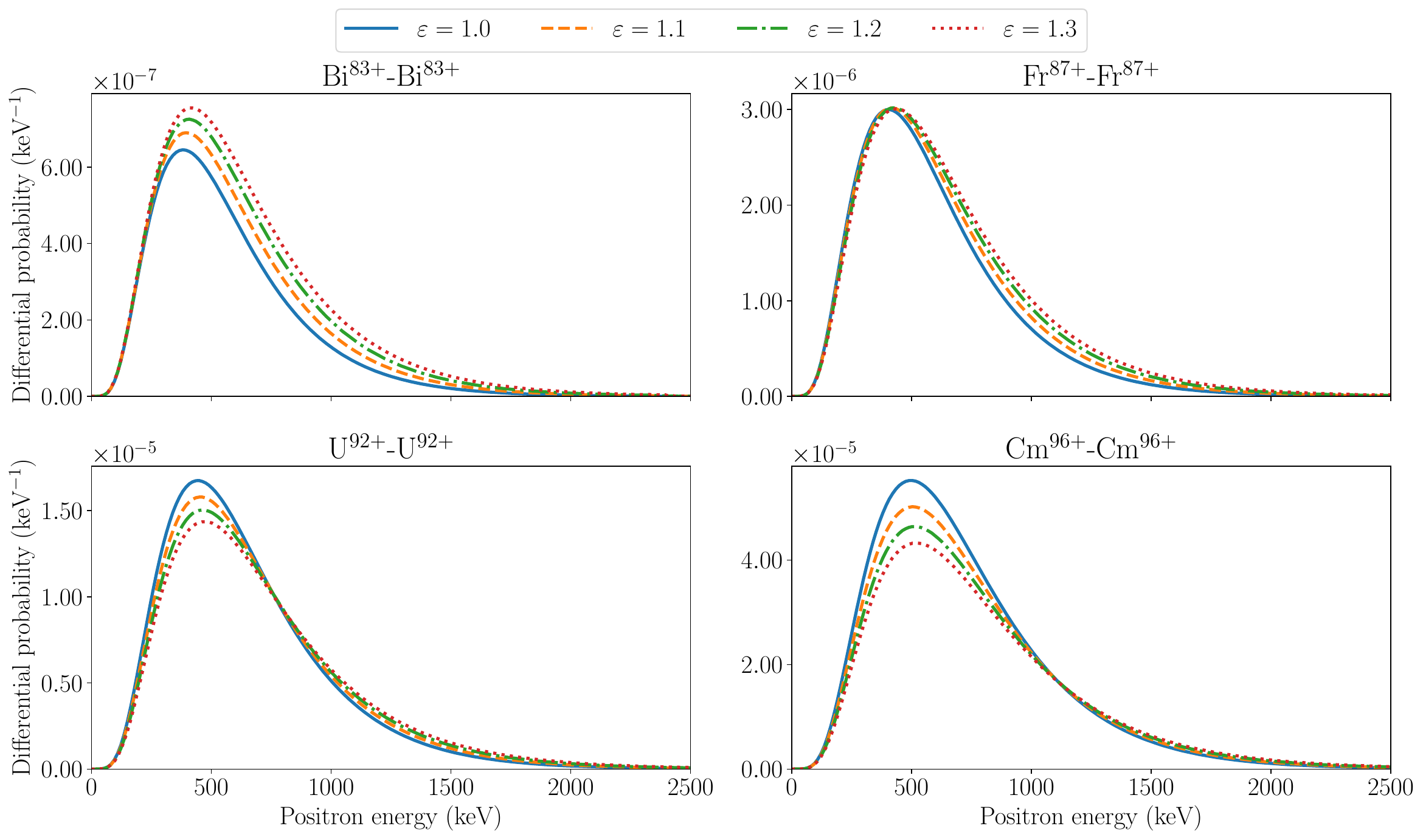}
    \caption{Energy spectra of positrons for symmetric collisions of nuclei with $Z=83$, $87$, $92$, $96$ and $\varepsilon=1.0$, $1.1$, $1.2$, $1.3$ at $R_{\mathrm{min}}=17.5$ fm.}
    \label{fig:en_state=tot}
\end{figure*}
The nuclear trajectories corresponding to different impact parameters differ by the scaled collision energy $\varepsilon$. The latter is defined as a ratio of the projectile energy $E$ in the frame where the target is initially at rest, for the trajectory under consideration, and the energy $E_{0}$ corresponding to the head-on collision with the same $R_{\mathrm{min}}$:
\begin{equation}
 \varepsilon = \frac{E}{E_{0}}.
\end{equation}
Since the projectile energies $E$ for collisions with fixed $R_{\mathrm{min}}$ and non-zero impact parameters are greater than $E_{0}$, the scaled collision energy $\varepsilon$ satisfies the inequality
\begin{equation}
 \varepsilon \ge 1.
\end{equation}
In Table~\ref{tab1}, we list the total positron creation probabilities in collisions with various nuclear charges and scaled collision energies. The total probabilities are calculated as a sum of contributions from the discrete quasimolecular states $1s_{1/2}$, $2p_{1/2}$, $2s_{1/2}$, $3p_{1/2}$, and $3s_{1/2}$. As one can see, our results are in good agreement with those of Popov et al.~\cite{Popov_2023_Spontaneous} for $Z=92$ and $Z=96$, where the data for comparison are available. The data of Ref.~\cite{Popov_2023_Spontaneous} were obtained beyond the monopole approximation using a one-center expansion of the nuclear potential over spherical harmonics. Generally, the positron creation probabilities in Ref.~\cite{Popov_2023_Spontaneous} are slightly larger than the present results, because they include transitions to the upper positron continuum from a larger number of positron vacuum discrete states as well as from the lower continuum. As one can see in Table~\ref{tab1}, for subcritical systems with $Z=83$ and $Z=87$, the total positron creation probabilities increase with increasing the scaled collision energy $\varepsilon$. This is well understood since the mechanism of positron creation in these systems is dynamical. A higher collision energy, hence a higher projectile velocity, favors transitions when an energy gap exists between the initial and final states. For the supercritical system with $Z=96$, the picture is totally different. In the supercritical regime, there is no energy gap between the initial and final positron states; transitions may occur with non-zero rate even at zero projectile velocity, and actually a lower collision energy favors positron creation, since the smaller the projectile velocity, the larger the time spent by the system in the supercritical regime, which results in a larger positron creation probability. The system with $Z=92$ thus represents an intermediate case: the total positron creation probabilities change only slightly with increasing the scaled collision energy  $\varepsilon$. Our data indicate a weak increase of the total positron-creation probability at $\varepsilon\rightarrow 1$ while in Ref.~\cite{Popov_2023_Spontaneous} it is close to a constant. The explanation can be as follows. For the collisions under consideration in the system with $Z=92$, the positron creation from the $1s_{1/2}$ state definitely has a large contribution from the spontaneous mechanism, since this quasimolecular state enters the upper positron continuum at $R\approx 35$~fm. Positron creation from the other vacuum positron states, however, is due to the dynamical mechanism. We retain only five positron vacuum states in the calculations, and the total probability is still dominated by the spontaneous contribution from the $1s_{1/2}$ state. Adding more vacuum states with the dynamical positron creation mechanism may slightly change the situation, as can be seen in the data of Ref.~\cite{Popov_2023_Spontaneous}.

\subsection{Angle-integrated energy distributions of outgoing positrons}
The energy spectra of outgoing positrons are obtained by integration of Eq.~(\ref{eq:distr}) over the emission angles. The energy spectra can be a more sensitive tool in detecting the spontaneous mechanism of the positron creation, compared with the total positron creation probabilities. 
A signature of the transition to the supercritical regime revealed in Refs.~\cite{Popov_2020_How,Popov_2023_Spontaneous} concerns the maximum in the energy distribution of emitted positrons. It was shown within the monopole approximation \cite{Popov_2020_How} and beyond \cite{Popov_2023_Spontaneous} that in the subcritical regime ($2Z<Z_{\mathrm{cr}}$) the maximum in the energy distribution increases with increasing $\varepsilon$, whereas in the supercritical regime ($2Z>Z_{\mathrm{cr}}$) the tendency is reversed, and the maximum in the energy distribution decreases with increasing $\varepsilon$.
Both the methods \cite{Popov_2020_How} and \cite{Popov_2023_Spontaneous} are based on a one-center representation of the positron wave functions, however. In the present work we examine the positron energy spectra obtained within a more realistic two-center description of the systems under consideration.

\begin{figure}
 \includegraphics[width=\columnwidth]{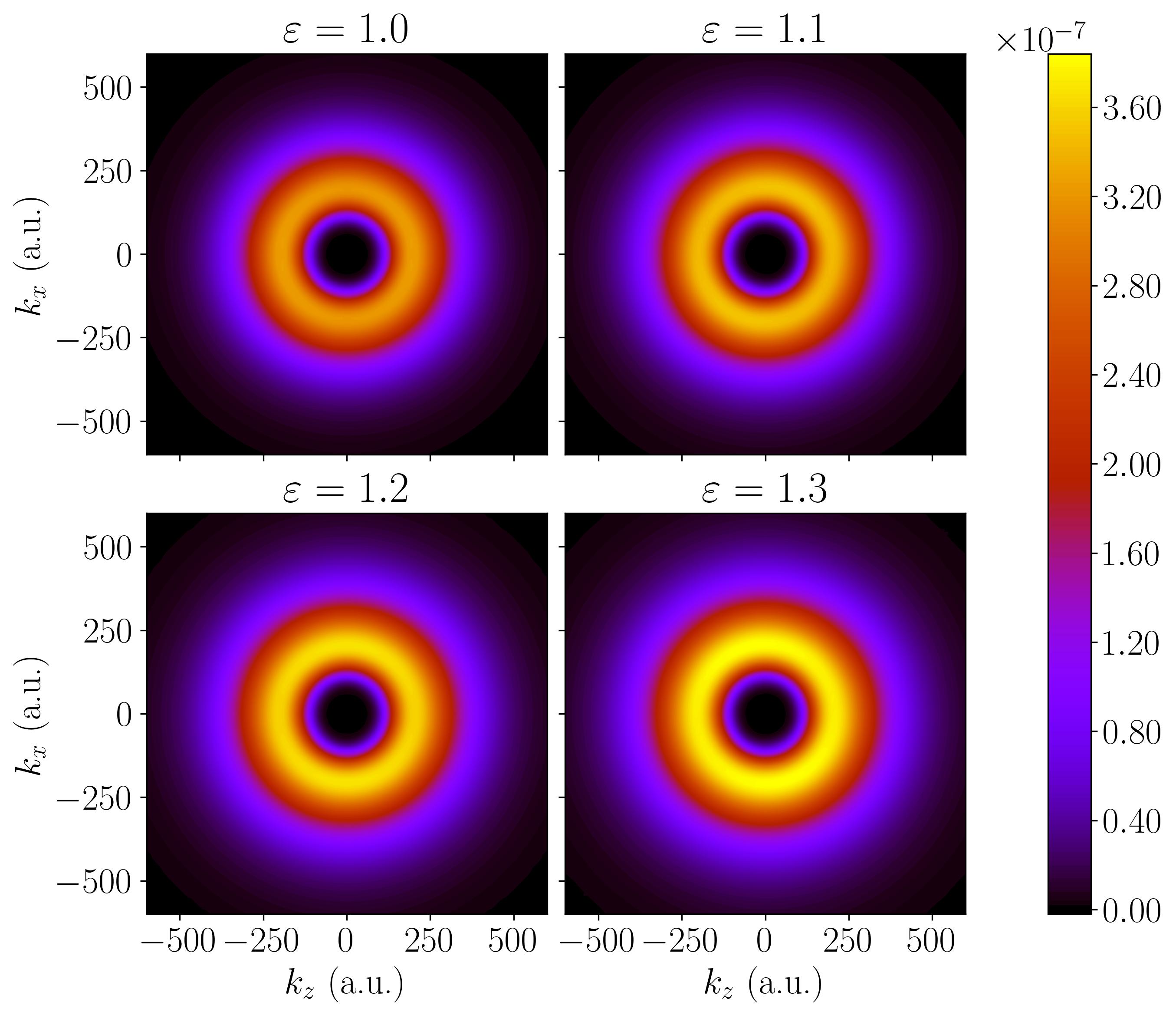}
        \caption{Pseudocolor plot of differential probabilities  (\ref{eq:distr}) corresponding to the emitted positron momentum projections $k_{z}$ and $k_{x}$ in the collision plane. Symmetric collision with $Z=83$, $R_{\mathrm{min}}=17.5$~fm, and $\varepsilon=1.0$, $1.1$, $1.2$, $1.3$. The color scale is linear.}
         \label{fig:enan_dens_one_plt_z=83_state}
\end{figure}
\begin{figure}
 \includegraphics[width=\columnwidth]{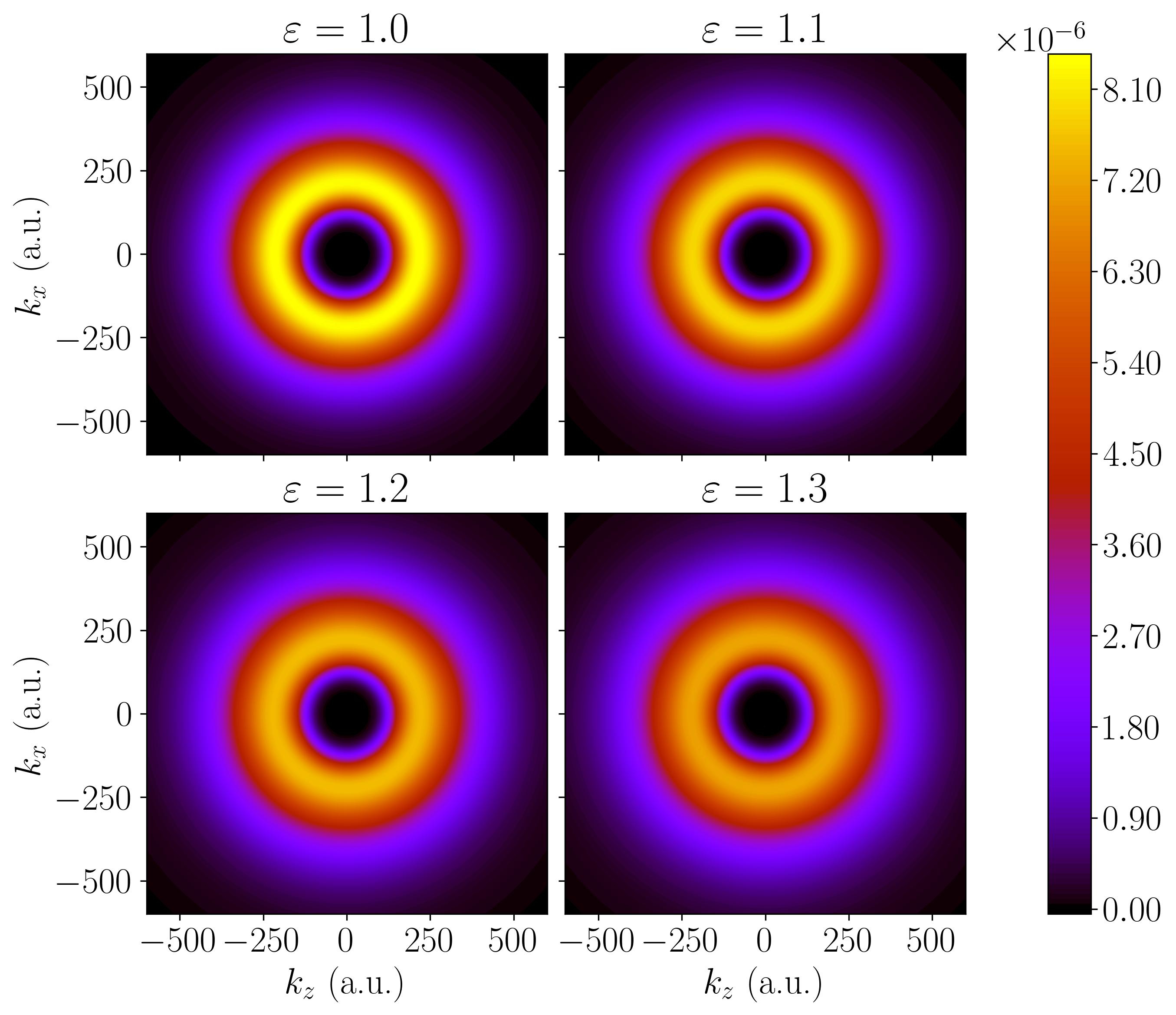}
        \caption{Pseudocolor plot of differential probabilities  (\ref{eq:distr}) corresponding to the emitted positron momentum projections $k_{z}$ and $k_{x}$ in the collision plane. Symmetric collision with $Z=92$, $R_{\mathrm{min}}=17.5$~fm, and $\varepsilon=1.0$, $1.1$, $1.2$, $1.3$. The color scale is linear.}
        \label{fig:enan_dens_one_plt_z=92_state}
\end{figure}
In Fig. \ref{fig:en_state=tot}, the positron energy spectra are presented for collisions of two identical nuclei with the charge numbers $Z=83$, $87$, $92$, $96$ and scaled energies $\varepsilon=1.0$, $1.1$, $1.2$, $1.3$. In the subcritical system, $Z=83$, the largest differential probability at the maximum of the distribution corresponds to $\varepsilon=1.3$; it gradually decreases as the scaled collision energy  changes from $1.3$ to $1.0$. For $Z=87$, the differential probability at the maximum of the distribution does not change as the scaled collision energy varies. Finally, in the supercritical systems with $Z=92$ and $Z=96$, where the total charge of the nuclei exceeds the critical value $Z_{\mathrm{cr}} \approx 175$ for $R_{\mathrm{min}}=17.5$~fm, the largest differential probability at the maximum of the distribution corresponds to $\varepsilon=1.0$ and gradually decreases as $\varepsilon$ varies from $1.0$ to $1.3$. These observations agree well with the previous results \cite{Popov_2020_How, Popov_2023_Spontaneous} obtained within the one-center approach. They also follow the trend seen in Table~\ref{tab1} for the total positron creation probabilities. A criterion based on the analysis of the positron energy spectra in the vicinity of the maximum of the energy distribution, however, seems to be more obvious. This can be explained as follows. Positrons created by the spontaneous mechanism cannot have their energies larger than the highest position of the supercritical resonance in the upper positron continuum during the collision. For the parameters used in the present calculations, this value does not exceed $550$~keV for $Z=96$ and $300$~keV for $Z=92$ but should be increased due to the resonance width. Therefore a broad tail of the positron energy spectrum is almost entirely due to the dynamical mechanism. For $Z=92$, the share of dynamically created positrons is large and masks the spontaneously created positrons in the total probabilities. On the other hand, when an analysis is performed on the differential probabilities in the vicinity of the maximum only, where the contribution of spontaneously created positrons is considerable, it reveals the supercritical regime of the positron production. For $Z=96$, the spontaneous mechanism dominates both the differential probability near the maximum and the total probability. 

\subsection{Energy and angular distributions of outgoing positrons}
Energy--angle distributions of outgoing positrons are presented in Figs.~\ref{fig:enan_dens_one_plt_z=83_state} and  \ref{fig:enan_dens_one_plt_z=92_state} for collisions in the systems with $Z=83$ and $Z=92$, respectively, for the scaled collision energies $\varepsilon$ equal to $1.0$, $1.1$, $1.2$, and $1.3$. Shown are the differential probabilities (\ref{eq:distr}) in the collision plane $z-x$ corresponding to the positron momentum projections $k_{z}$ and $k_{x}$, which are related to the positron kinetic energy and emission angles as follows:
\begin{equation}
\begin{split}
E_{k}&=c^{2}\left[\sqrt{1+\frac{k_{z}^{2}+k_{x}^{2}}{c^{2}}}-1\right],\\ \vartheta_{k}&=\arccos\left(\frac{k_{z}}{k_{z}^{2}+k_{x}^{2}}\right),\\
\varphi_{k}&=0\quad (k_{x}>0)\quad\mathrm{or}\quad\varphi_{k}=\pi\quad (k_{x}<0).
\end{split}
\end{equation}

Two main observations can be drawn from Figs.~\ref{fig:enan_dens_one_plt_z=83_state} and  \ref{fig:enan_dens_one_plt_z=92_state}. First, clearly the distributions do not exhibit any considerable anisotropy in both subcritical and supercritical regimes. Actually, they are nearly spherically symmetric for all the nuclear charges $Z$ and scaled energies $\varepsilon$ used in the calculations. This is understandable since creation of the electron-positron pairs mainly occurs on very short internuclear separations where the shape of the quasimolecule is close to the united atom with the spherical symmetry. Second, the same signature of transition from the subcritical regime to the supercritical regime as in the energy spectra can be seen in the energy--angle distributions. For $Z=83$ (Fig.~\ref{fig:enan_dens_one_plt_z=83_state}), with increasing $\varepsilon$, the bright ring corresponding to the maximum in the energy spectrum becomes even brighter, that is the differential probability at the maximum increases with increasing $\varepsilon$. On the contrary, for $Z=92$ (Fig.~\ref{fig:enan_dens_one_plt_z=92_state}), the bright ring corresponding to the maximum in the energy spectrum becomes dimmer as $\varepsilon$ increases, that is the differential probability at the maximum decreases.

\begin{figure}
\includegraphics[width=\columnwidth]{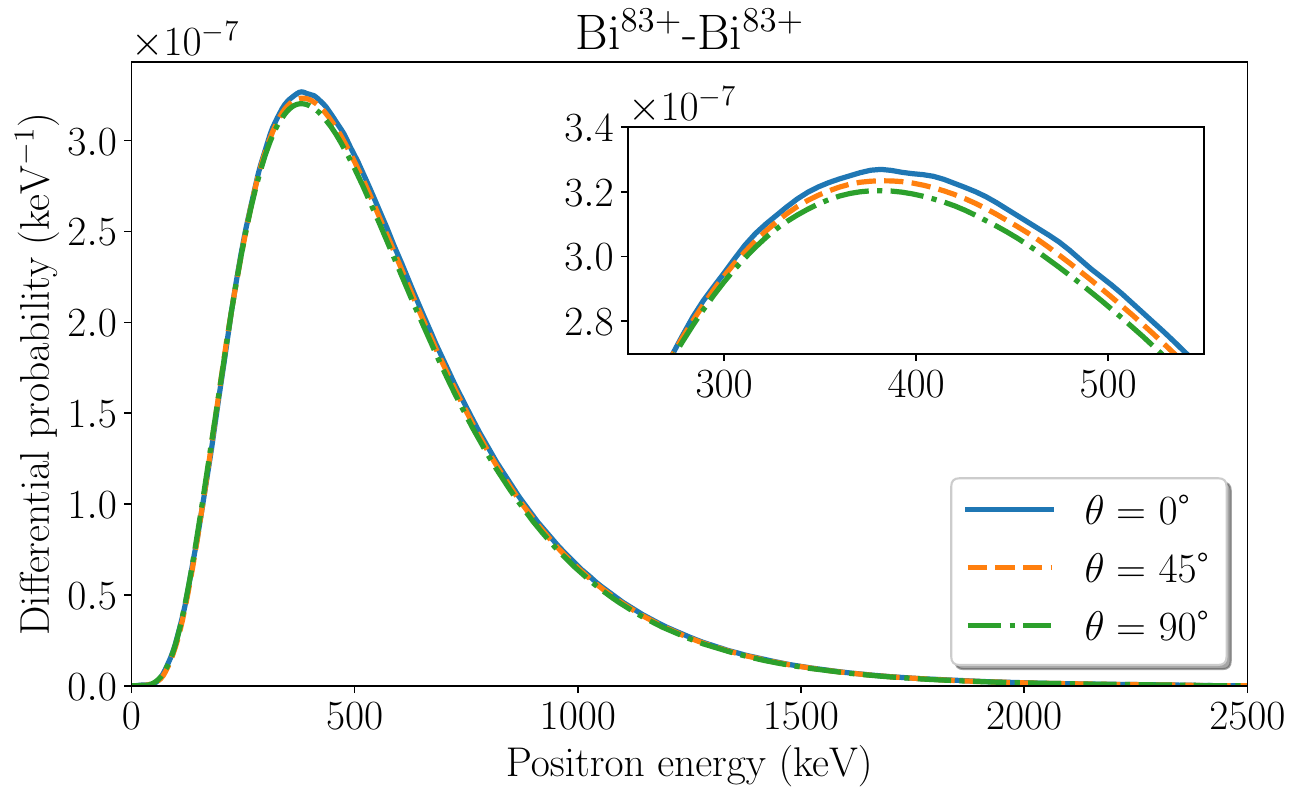}
        \caption{Energy spectra of emitted positrons for head-on symmetric collisions of nuclei with $Z=83$ at $R_{\mathrm{min}}=17.5$ fm for the angle $\theta=0^\circ$, $45^\circ$ and $90^\circ$ between the $z$-axis in the initial reference frame and the observation direction.}
         \label{fig:en_in_all_an_z=83_state=tot}
\end{figure}
\begin{figure}
        \includegraphics[width=\columnwidth]{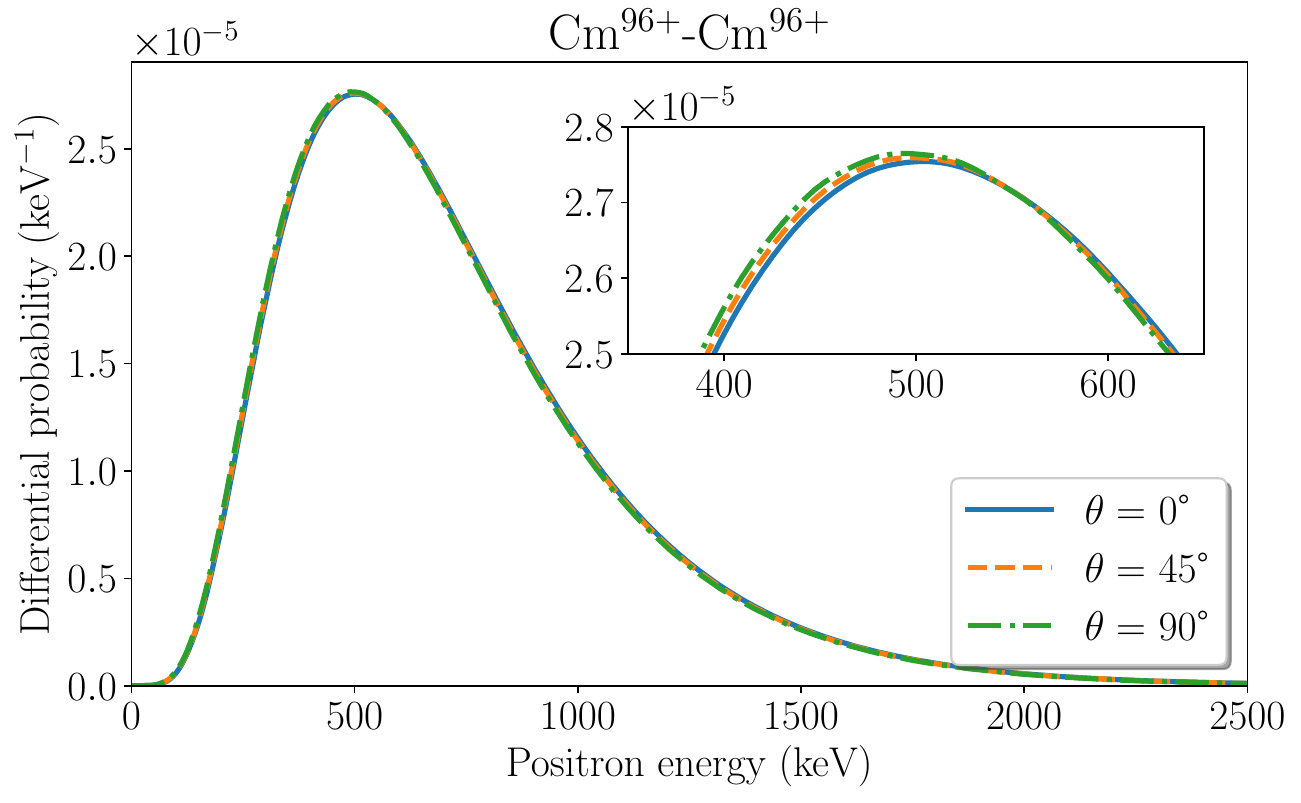}
        \caption{Energy spectra of emitted positrons for head-on symmetric collisions of nuclei with $Z=96$ at $R_{\mathrm{min}}=17.5$ fm for the angle $\theta=0^\circ$, $45^\circ$ and $90^\circ$ between the $z$-axis in the initial reference frame and the observation direction.}
        \label{fig:en_in_all_an_z=96_state=tot}
\end{figure}

An anisotropy in the distributions of outgoing positrons, although very small, still can be detected in our calculations. It is better seen in the angle-resolved energy spectra, presented for the positron emission  angles $\theta_{k}=0^\circ$, $\theta_{k}=45^\circ$, and $\theta_{k}=90^\circ$ in Figs.~\ref{fig:en_in_all_an_z=83_state=tot} and \ref{fig:en_in_all_an_z=96_state=tot} for symmetric head-on collisions ($\varepsilon=1$) with $Z=83$ and $Z=96$, respectively. It appears that the distributions are more anisotropic in the subcritical regime. For $Z=83$, 
the maximum of the distribution in Fig.~\ref{fig:en_in_all_an_z=83_state=tot} is highest at $\theta=0^\circ$ and lowest at $\theta=90^\circ$. The relative anisotropy parameter can be defined as a difference of these two values divided by the value at $\theta=0^\circ$. For $Z=83$, it is equal to $0.02$. The anisotropy is much less pronounced for $Z=96$ in Fig.~\ref{fig:en_in_all_an_z=96_state=tot}, the relative anisotropy parameter in this case is equal to $0.005$. A possible cause for this difference can be as follows. For collisions with $2Z<Z_{\mathrm{cr}}$ in the subcritical regime, the spontaneous channel is closed and all positrons are created through the dynamic mechanism. Dynamic pair creation occurs on larger internuclear distances as well, where the two-center geometry of the quasimolecule is well shaped and can affect the angular distributions of outgoing positrons. In contrast, for collisions with $2Z>Z_{\mathrm{cr}}$, the  spontaneous channel is open if the internuclear distance is less than the critical value. The larger the nuclear charge $Z$ in the supercritical regime, the more significant the contribution of the spontaneous mechanism. Since the spontaneous vacuum decay occurs when the nuclei are in close proximity to each other (for $Z=96$, the critical internuclear distance is $52$~fm), the quasimolecule is close to the united atom limit, and angular distributions of outgoing positrons are almost isotropic. In this respect, we should note that our calculations do not confirm a prediction of considerable anisotropy in the angular distributions of positrons \cite{Popov_1979}, which was based on an analytical semiclassical approach.

\section{Conclusion}
In this paper, we have studied electron-positron pair production in slow collisions of two identical heavy nuclei. Our theoretical approach is based on the time-dependent Dirac equation for a positron moving in the field of two nuclei. This approach is essentially two-center and beyond the monopole approximation. The computations are done with the help of the generalized pseudospectral method in modified prolate spheroidal coordinates, which provides an accurate and reliable treatment of two-center quantum systems up to very small distances between the centers.
We have calculated the total positron creation probabilities as well as distributions of the outgoing positrons with respect to the energy and emission angles. The distributions are calculated by projecting the wave packet of outgoing positrons onto the plane waves in the spatial region far from the nuclei, so the influence of the Coulomb field from the nuclei is negligible, and plane-wave approximation for the final positron states is well justified. The results are obtained for both subcritical and supercritical collisions, when a channel of spontaneous positron creation opens. The signatures of transition to the supercritical regime, which were previously revealed in the angle-integrated energy spectra of positrons \cite{Popov_2020_How, Popov_2023_Spontaneous}, have been confirmed in the angle-resolved distributions. The angular distributions of outgoing positrons appear almost isotropic, and this is not surprising since the positron creation occurs mostly at very short internuclear distances where the quasimolecule is close to the united atom limit with the spherical symmetry. We should note, however, that when solving the time-dependent Dirac equation in the rotating frame of reference, we have neglected the rotational coupling term. Previously it was discussed in the literature that the influence of this term on the total probabilities of positron creation in slow collisions is negligible. A question how the rotational coupling affects the energy and angular distributions of outgoing positrons remains open and requires further research.

\begin{acknowledgments}
This work was supported by the Russian Science Foundation (Grant No 22-62-00004, https://rscf.ru/project/22-62-00004/).
\end{acknowledgments}

\bibliography{main.bib}
\end{document}